# CAPÍTULO 4

## ALGORITMO PARA GERAÇÃO DE CONTORNO DE MALHAS RETANGULARES PARA CÁLCULO DE DIFERENÇAS FINITAS

**Pedro Zaffalon da Silva**
Universidade Estadual de Londrina
Paraná - Londrina
http://lattes.cnpq.br/6079988422025271

**Neyva Maria Lopes Romeiro**
Universidade Estadual de Londrina
Paraná - Londrina
http://orcid.org/0000-0002-3249-3490

**Rafael Furlanetto Casamaximo**
Universidade Estadual de Londrina
Paraná - Londrina
http://lattes.cnpq.br/5123297463582913

**Iury Pereira de Souza**
Universidade Estadual de Londrina
Paraná - Londrina
http://lattes.cnpq.br/2572023393070173

**Paulo Laerte Natti**
Universidade Estadual de Londrina
Paraná - Londrina
http://orcid.org/0000-0002-5988-2621

**Eliandro Rodrigues Cirilo**
Universidade Estadual de Londrina
Paraná - Londrina
http://orcid.org/0000-0001-7530-1770

**RESUMO**: Neste trabalho, um algoritmo para geração de contorno 2D envolvendo regiões irregulares é proposto. Nesse algoritmo, o contorno do domínio físico é aproximado para segmentos de malha utilizando as coordenadas do contorno dado. Para este propósito, o código usa uma estrutura de repetição que analisa as coordenadas do contorno irregular conhecidas para aproximar o contorno do domínio físico para segmentos de malha. Para isso, calcula-se o coeficiente linear da reta definida pelos pontos do contorno conhecido e seus vértices vizinhos. Desta forma, o algoritmo calcula os pontos da linha e sua distância para os nós da malha mais próximos, permitindo obter pontos do contorno aproximado. Esse processo é repetido até que o contorno aproximado completo seja gerado. Sendo assim, um algoritmo para geração de contorno aproximado, sob os nós da malha, torna-se adequado para descrever malhas envolvendo geometrias com contornos irregulares para fins de cálculo de diferenças finitas, resultando em simulações numéricas da modelagem matemática de fenômenos naturais. O algoritmo é analisado usando três geometrias, onde avalia-se a diferença entre as áreas delimitadas pelo contorno dado e aproximado em porcentagens, número de nós e a quantidade de pontos internos. Verifica-se que quanto maior a complexidade da geometria, maior a quantidade de nós no contorno é necessária, exigido desta forma, malhas mais refinadas, para obter diferenças de áreas abaixo de 2%.
**PALAVRAS-CHAVE:** Geometria irregular, contorno, malha, diferenças finitas.

### RECTANGULAR MESH CONTOUR GENERATION ALGORITHM FOR FINITE DIFFERENCES CALCULUS
**ABSTRACT:** In this work, a 2D contour generation algorithm, involving irregular regions is proposed.

In this algorithm, the contour of the physical domain is approximated by mesh segments using the coordinates of the known contours. For this purpose, the algorithm uses one repetition structure that analyzes the coordinates of the known irregular contours to approximate the contour of the physical domain by mesh segments. To this end, the algorithm calculates the slope of the line defined by the known point of the irregular contours and the neighboring vertices. Then the algorithm calculates the line points and the shortest distance from these points to a mesh node, thus generating a point of the approximate contour. This process is repeated until the approximate contour is obtained. Therefore, an approximate contour generation algorithm, under mesh nodes, becomes appropriate to describe irregular contours geometries used in finite difference method, allowing numerical simulations of mathematical modelling of natural phenomena. The algorithm is analyzed using tree geometries, which are evaluated by the difference between the area bounded by the known and approximate contour, the number of nodes on the contour and inside the geometry. It is observed that the more complex the geometry, the more nodes are necessary in the contour, demanding more refined meshes, to obtain area differences below 2%.

## INTRODUÇÃO

A modelagem e simulação, por meio da manipulação de equações diferenciais, constitui uma importante ferramenta para a análise e a descrição matemática de diversos fenômenos. Entretanto, devido ao fato de que, a grande maioria das equações diferenciais não possui solução analítica, faz-se necessário a aplicação de métodos numéricos para sua resolução.

Para aplicar métodos numéricos, é necessário conhecer informações da geometria do meio que está sendo investigado, considerando a malha computacional. A malha computacional consiste na representação discretizada do domínio físico descrito através de um contorno dado. Assim, a malha é formada por um conjunto de células, limitadas pelas arestas, nos quais são denominadas de faces, contendo vértices, que são chamados de nós.

No entanto, na modelagem de fenômenos naturais, raramente o domínio onde estão definidas as condições de contorno do problema, encontra-se sob os nós da malha computacional [3]. Assim, malhas cartesianas em num plano bidimensional, se deparam com sérias dificuldades ao prescrever condições de contorno em domínios não regulares, dificultando a resolução do problema considerando o método de diferenças finitas [12]. Porém, discretizações utilizando malhas cartesianas são atraentes em termos de eficiência e baixo armazenamento de memória [5].

Neste contexto, muitos autores empregam métodos que utilizam interpolações polinomiais algébricas para construir as equações de diferenças nos pontos do contorno dado, tornando-se possível incorporar o contorno irregular ao método, ou seja todos os cálculos sobre domínios irregulares reduzem-se aos domínios regulares, obtendo assim uma solução numérica mais precisa para o problema [2, 7, 8, 12].

Desta forma, este trabalho apresenta o desenvolvimento de um algoritmo, utilizando *software Octave* [4], que descreve adequadamente o contorno de uma região irregular por meio de um conjunto finito de pontos, resultando em um domínio mapeado. Através do algoritmo, torna-se possível utilizar o método diferenças finitas para resolver numericamente equações diferenciais parciais utilizando malhas que contém contornos irregulares.

O procedimento utilizado emprega a técnica que consiste em representar retas definidas pelos pontos do contorno dado, para gerar os pontos do contorno aproximado mais próximos dos pontos de malha. Inicialmente define-se como as coordenadas dos pontos do contorno dado se encontram ordenadas, na sequência apresenta-se dois casos onde o algoritmo verifica se é necessário excluir nós externos em convexidades ou adicionar nós em concavidades. Após este desenvolvimento, apresenta-se um resumo do algoritmo para obter o contorno aproximado. Como verificação, apresenta-se, nos resultados, malhas geradas pelo algoritmo. Finalmente, as conclusões são apresentadas.

**DESENVOLVIMEMTO**

Sendo o contorno dado, inscrito em uma região de domínio retangular $R = [X_0, X_f] \times [Y_0, Y_f]$, define-se $\delta_x = \frac{X_f - X_0}{N_x}$ e $\delta_y = \frac{Y_f - Y_0}{N_y}$, onde $N_x$ e $N_y$ são as partições em *x* e *y*, respectivamente. A partir desses valores, define-se a malha na qual o domínio irregular será representado, por meio do contorno aproximado.

Para obter o contorno aproximado, foi considerado um algoritmo que utiliza funções lineares, que representa retas definidas pelos pontos do contorno dado, para gerar o contorno aproximado mais próximos dos pontos da malha. Desta forma, o algoritmo recebe como parâmetros de entrada o vetor com as coordenadas *x* e *y* do contorno dado, os valores mínimos das coordenadas deste vetor, representado por ($x_{min}$, $y_{min}$), e o espaçamento da malha, $\delta_x$ e $\delta_y$.

Também, é definido o sentido no qual os pontos do contorno dado encontram-se ordenados. Assim, o primeiro ponto do contorno aproximado, ou como denotado, primeiro nó, será utilizado como base para calcular os demais nós. Este nó será obtido aproximando o ponto inicial do contorno dado a um nó mais próximo, interno a região da área do contorno aproximado, sob um ponto de malha. Entretanto, para isso faz-se necessário verificar em qual direção a região interna, da área de interesse, se encontra. Para verificar está região, deve-se observar qual o sentido em que as coordenadas dos pontos do contorno estão ordenadas, se é horário ou anti-horário. Para definir como encontra-se ordenado o contorno, utiliza-se as Figuras 1a-d.

Na Figura 1a, as coordenadas encontram-se ordenadas no sentido horário, onde observa-se que em cada parte dos segmentos de retas de cor azul, o ponto subsequente apresenta valor maior para a coordenada *x*, e a área interna da figura localiza-se abaixo dos segmentos de retas. Por outro lado, nos segmentos de reta de cor vermelha, o ponto subsequente apresenta valor menor para a coordenada *x*, e a área interna encontra-se acima dos segmentos de retas. Na Figura 1b, as coordenadas encontram-se ordenadas no sentido anti-horário. Observa-se que em cada parte dos segmentos de retas de cor azul, o ponto subsequente apresenta valor maior para a coordenada *x*, e a área interna da figura localiza-se acima dos segmentos de retas. Nos segmentos de reta de cor vermelha, o ponto subsequente apresenta valor menor para a coordenada *x* e a área interna encontra-se abaixo dos segmentos de retas.

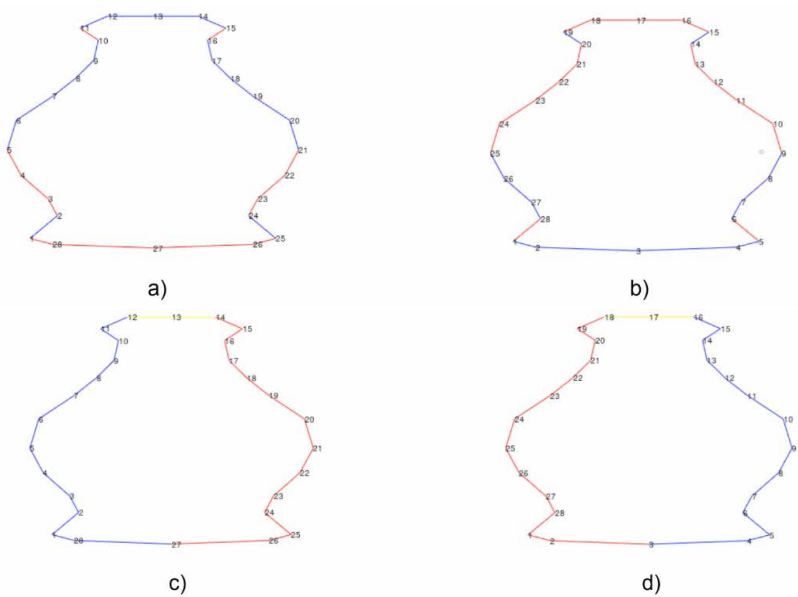

Figura 1. Os pontos do contorno encontram-se ordenados nos sentidos: a) e c) horário; b) e d) anti-horário.

Fonte: Os autores.

Na Figura 1c, as coordenadas encontram-se ordenadas no sentido horário. observa-se que nos segmentos de retas de cor azul, o ponto subsequente apresenta valor maior para *y*, e a área interna da figura localiza-se à direita do ponto. Por outro lado, nos segmentos de retas de cor vermelha o ponto subsequente apresenta valor menor de *y*, e a área interna encontra-se à esquerda. Ainda, nesta figura, pode-se observar segmentos de retas na cor amarela, onde não ocorre variação no valor de *y*, neste caso a direção será a mesma do ponto anterior. Na Figura 1d, as coordenadas encontram-se ordenadas no sentido anti-horário. observa-se que nos segmentos de retas de cor azul, o ponto subsequente apresenta valor maior para *y*, e a área interna da figura localiza-se à esquerda do ponto. Por outro lado, nos segmentos de retas de cor vermelha o ponto subsequente apresenta valor menor de *y*, e a área interna está à direita. Similarmente a Figura 1c, nos segmentos de retas na cor amarela, não ocorre variação no valor de *y*, neste caso a direção será a mesma do ponto anterior.

Verificado qual a direção em que a área interna da região se encontra, o algoritmo aproximar o primeiro ponto do contorno dado para o nó interno mais próximo. Após obter o nó inicial, o algoritmo calcula as coordenadas dos nós aproximados até o segundo ponto do contorno dado, conforme exemplificado nas Figuras 2a-d.

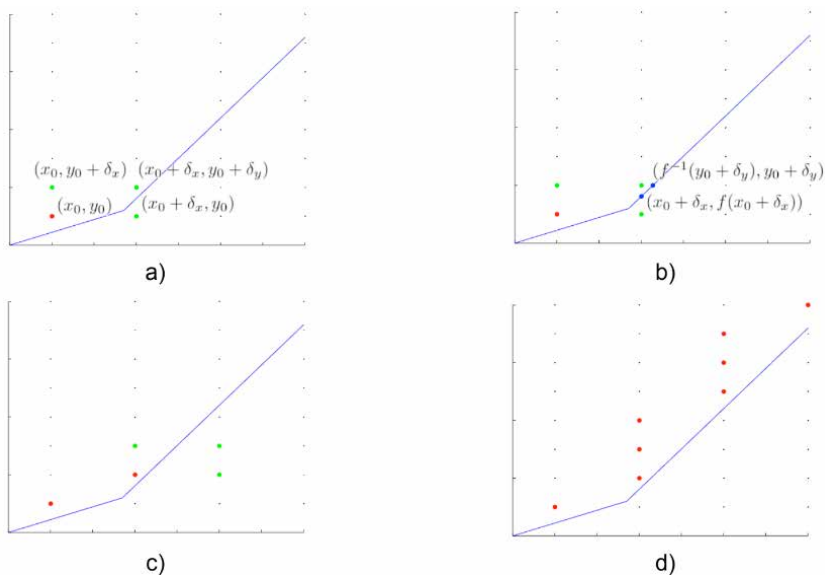

Figura 2. Etapas do algoritmo para obter os nós entre dois pontos.

Fonte: Os autores.

Na Figura 2a, o ponto ($x_0$, $y_0$) em vermelho, representa o primeiro ponto do contorno aproximado, as retas azuis descrevem o contorno dado no sentido anti-horário. Devido a reta ser crescente em *x* e em *y*, o próximo nó do contorno aproximado deve ter coordenadas maior ou igual, tanto em *x* quanto em *y*. Sendo assim, tem-se três possibilidades para o próximo nó, sendo ($x_0 + \delta_x$, $y_0$), ($x_0$, $y_0 + \delta_x$) e ($x_0 + \delta_x + y_0 + \delta_y$), como pode ser identificado pela cor verde na Figura 2a.

Com o objetivo de identificar qual nó deve ser adicionado ao contorno aproximado, é necessário verificar a distância de cada ponto em relação ao contorno dado. Para isso, utiliza-se a função $f(x)$, que representa a reta definida pelo ponto atual do contorno dado e seu antecessor. Obtido a função, calcula-se os valores de $f(x_0 + \delta_x)$ e de $g(y_0 + \delta_y)$, sendo $g(x) = f^{-1}(x)$, os quais permitem obter a distância em relação aos nós analisados.

Para identificar qual nó deve ser adicionado, são analisados dois pontos na reta, sendo $X = x_0 + \delta_x$ e $Y = y_0 + \delta_y$. Desta forma, obtêm-se ($x_0 + \delta_x$, $f(x_0 + \delta_x)$) e $f^{(-1)}(y_0 + \delta_y)$, $y_0 + \delta_y$), conforme ilustrado na Figura 2b. Verifica-se que o ponto que deve ser adicionado no contorno aproximado é ($x_0 + \delta_x$, $y_0 + \delta_y$) pois a distância entre a reta e o ponto ($x_0$, $y_0 + \delta_y$) é maior que $\delta_x$ e ($x_0 + \delta_x$, $y_0$) não é um ponto interno à figura. Obtido o segundo nó, continua-se o processo realizando as mesmas operações, conforme ilustrado na Figura 2c. Interrompe-se a etapa ao obter o nó mais próximo ao ponto atual do contorno dado, conforme ilustrado na Figura 2d. Finalizada a etapa para o ponto atual, o mesmo procedimento será realizado para os demais pontos do contorno dado, de forma que, ao passar por todos os pontos, o contorno aproximado é obtido.

## EXCEÇÕES: CONVEXIDADES E CONCAVIDADES

Em certos casos, nos quais o sentido de ou de muda em relação à reta anterior, é necessário realizar verificações para gerar o contorno aproximado correto. Assim, tem-se aos casos 1 e 2.

**Caso 1**: Em convexidades os nós obtidos podem ser externos em relação à próxima reta, como ilustra a Figura 3.

Na Figura 3 o ponto *P* é o nó mais próximo acima da primeira reta do contorno, porém em relação à reta seguinte está externa à figura. Com o objetivo de evitar esta situação, é realizado uma verificação para excluir nós externos em convexidades.

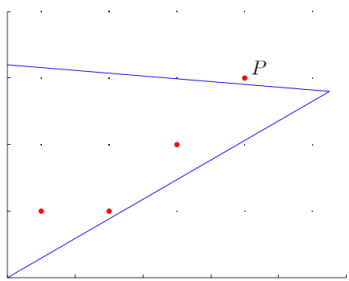

Figura 3. Exemplo de nó externo em relação à próxima reta.
Fonte: Os autores.

Na Figura 3 o ponto *P* é o nó mais próximo acima da primeira reta do contorno, porém em relação à reta seguinte está externa à figura. Com o objetivo de evitar esta situação, é realizado uma verificação para excluir nós externos em convexidades.

**Caso 2:** Em concavidades, o último nó obtido pode não ser o nó inicial da próxima reta, como descreve a Figura 4.

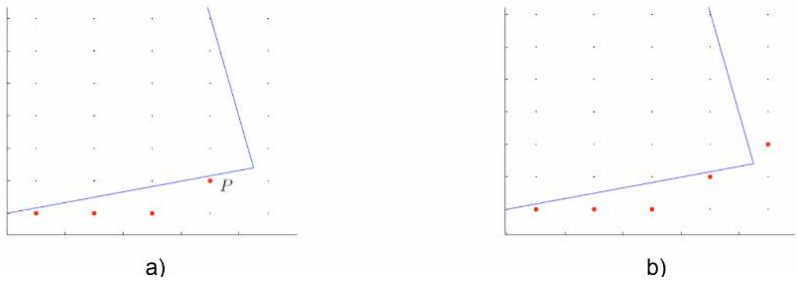

Figura 4. Exemplo de um caso no qual é necessário adicionar um nó adicional.

Fonte: Os autores.

Na Figura 4a, o ponto *P* é o último nó da primeira reta do contorno e, consequentemente, o inicial para a reta seguinte. Considerando que a reta possui valor crescente em *y* e decrescente em *x*, o próximo nó obtido seria externo. Isso ocorre porque o último nó obtido não é adequado como valor inicial para a reta seguinte. Para evitar nós incorretos, um novo nó é adicionado no contorno neste caso, realizando a mesma operação, porém verificando em relação à próxima reta, obtendo o nó inicial adequado, conforme ilustrado na Figura 4b.

### ALGORITMO

Todo o desenvolvimento descrito encontra-se apresentado de forma resumida no Algoritmo 1. Os parâmetros $v_x$ e $v_y$ representam a diferença entre dois pontos do contorno dado.

### RESULTADOS

Utilizando o Algoritmo 1, apresenta-se resultados de malhas retangulares, considerando as geometrias da mama, da garrafa e do avião. Na geometria da mama, os autores Foucher, Ibrahim e Saad [6] e Manganin *et al*. [9] apresentam simulações numéricas utilizando equações diferenciais parciais, para descreverem o crescimento tumoral, enquanto as geometrias da garrafa e do avião encontram-se ilustradas em Naozuka [10]. Naozuka [10] e Naozuka *et al* [11], utilizaram as geometrias para avaliar o gerador de malhas desenvolvido em coordenadas generalizadas, envolvendo técnicas de multiblocos. Também, devido ao procedimento utilizado, avaliou a qualidade dos elementos de malha.

Os pontos do contorno das geometrias foram coletados utilizando o programa *WebPlotDigitizer* 4.3 [13]. As malhas internas foram obtidas utilizando a função *inpolygon* do *Octave* [4], a partir do contorno aproximado das figuras. As áreas das geometrias foram obtidas pelo método de Gauss, onde calcula a área, , de polígonos irregulares a partir do conjunto de coordenadas dos vértices do polígono ordenados no sentido anti-horário [1], utilizando a equação (1)

$$A = \frac{1}{2}\left\{\begin{vmatrix} x_0 & x_1 \\ y_0 & y_1 \end{vmatrix} + \begin{vmatrix} x_1 & x_2 \\ y_1 & y_2 \end{vmatrix} + \ldots + \begin{vmatrix} x_{n-2} & x_{n-1} \\ y_{n-2} & y_{n-1} \end{vmatrix} + \begin{vmatrix} x_{n-1} & x_n \\ y_{n-1} & y_n \end{vmatrix}\right\} \quad (1)$$

onde $x_n$ e $y_n$, representam os pontos a serem utilizados para calcular a área tanto da região do domínio dado e do domínio aproximado.

Utilizando a equação (1), apresenta-se na Tabela 1, diferenças entre as áreas delimitadas pelo contorno dado e aproximado em porcentagens, das geometrias da mama, garrafa e avião, utilizando diferentes refinamentos de malha. Informa-se o número de nós do contorno dado, do contorno obtido nos refinamentos e a quantidade de nós internos, para fins de cálculo de diferenças finitas.

**Algoritmo 1:** Algoritmo para obter o contorno dado.

**Entrada:** vetor do contorno dado $x_{min}$, $y_{min}$, $\delta_x$ e $\delta_y$
**Saída:** Contorno Aproximada

1 **início**
2     Aproxime a primeira coordenada para o nó mais próximo da malha;
3     Adicione o nó obtido ao contorno aproximado
4     **para** $i \leftarrow 2$; $i <$ tamanho do contorno dado; $i$++ **faça**
5         Calcule a diferença entre o ponto atual e o anterior do contorno dado, tanto para $x(v_x)$ quanto para $y(v_y)$;
6         Calcule o número de nós entre o último nó e o ponto atual do contorno dado, tanto para $x(n_x)$ quanto para $y(n_y)$;
7         **se** $n_x \geq 1$ ou $n_y \geq 1$ **então**
8             **se** $v_x == 0$ **então**
9                 **para** $j \leftarrow 1$; $j < n_y$; $j$ ++ **faça**
10                     Adicione um novo nó no contorno aproximado, com coordenada igual ao anterior somada com $\delta_y$
11                 **fim**
12             **fim**
13             **senão, se** $v_y == 0$ **então**
14                 **para** $j \leftarrow 1$; $j < n_y$; $j$ ++ **faça**
15                     Adicione um novo nó no contorno aproximado, com coordenada igual ao anterior somada com $\delta_x$
16                 **fim**
17             **fim**
18             **senão**
19                 Calcule o coeficiente angular da reta definida pelo atual e anterior $(v_y \backslash v_x)$
20                 **enquanto** Diferença entre o último nó obtido e o ponto atual em $x \geq \delta_x$ ou diferença em $y \geq \delta_y$
21                     Calcule os 3 possíveis próximos nós;
22                     Verifique qual é o nó interno mais próximo da reta;
23                     Adicione o nó obtido ao contorno aproximado
24                 **fim**
25             **fim**
26         **Se** Região apresentar convexidade **então**
27             **enquanto** Último nó estiver externo à figura **faça**
28                 Remova o último nó;
29             **fim**
30         **fim**
31         **Senão se** Região apresentar convexidade **então**
32             Obter o próximo nó, conforme o método anterior
33         **fim**
34     **fim**
35 **fim**

Os refinamentos, definidos por $N_i$ e $N_j$, $i = 1,2,...,N_x$ e $j = 1,2,..., N_y$, descrevem os números de partições nas direções *x* e *y* respectivamente.

| $N_i = N_j$ | Diferença entre as áreas (%) | Nós no contorno aproximado | Nós internos |
|---|---|---|---|
| **Mama** - 84 nós no contorno dado | | | |
| 50 | 4,3922 | 164 | 2000 |
| 70 | 2,8735 | 232 | 3934 |
| 80 | 2,6452 | 265 | 5131 |
| 100 | 1,9865 | 332 | 8029 |
| 120 | 1,6248 | 398 | 11564 |
| 150 | 1,2510 | 501 | 18075 |
| 200 | 0,8677 | 665 | 32144 |
| 250 | 0,6287 | 835 | 50242 |
| 300 | 0,5073 | 1002 | 72336 |
| **Garrafa** - 30 nós no contorno dado | | | |
| 50 | 5,4397 | 164 | 2000 |
| 70 | 3,2493 | 232 | 3934 |
| 80 | 2,6453 | 265 | 5131 |
| 100 | 2,6869 | 332 | 8029 |
| 120 | 1,8939 | 398 | 11564 |
| 150 | 1,7630 | 501 | 18075 |
| 200 | 1,2857 | 665 | 32144 |
| 250 | 1,0239 | 835 | 50242 |
| 300 | 0,8344 | 1002 | 72336 |
| **Avião** - 230 nós no contorno dado | | | |
| 70 | 11,6311 | 294 | 1239 |
| 100 | 9,0347 | 430 | 2508 |
| 150 | 5,5745 | 649 | 5680 |
| 200 | 4,5470 | 871 | 10060 |
| 300 | 2,9122 | 1313 | 22683 |
| 350 | 2,5392 | 1533 | 30862 |
| 400 | 2,2523 | 1754 | 40301 |
| 450 | 1,9190 | 1975 | 51053 |
| 500 | 1,7194 | 2196 | 63033 |

Tabela 1. Resultados obtidos nas geometrias da mama.

Na Tabela 2 têm-se os resultados dos contornos dados e aproximados e, as malhas geradas pelo Algoritmo 1, considerando as geometrias da mama, garrafa e avião em quatro dos nove refinamentos dados na Tabelas 1.

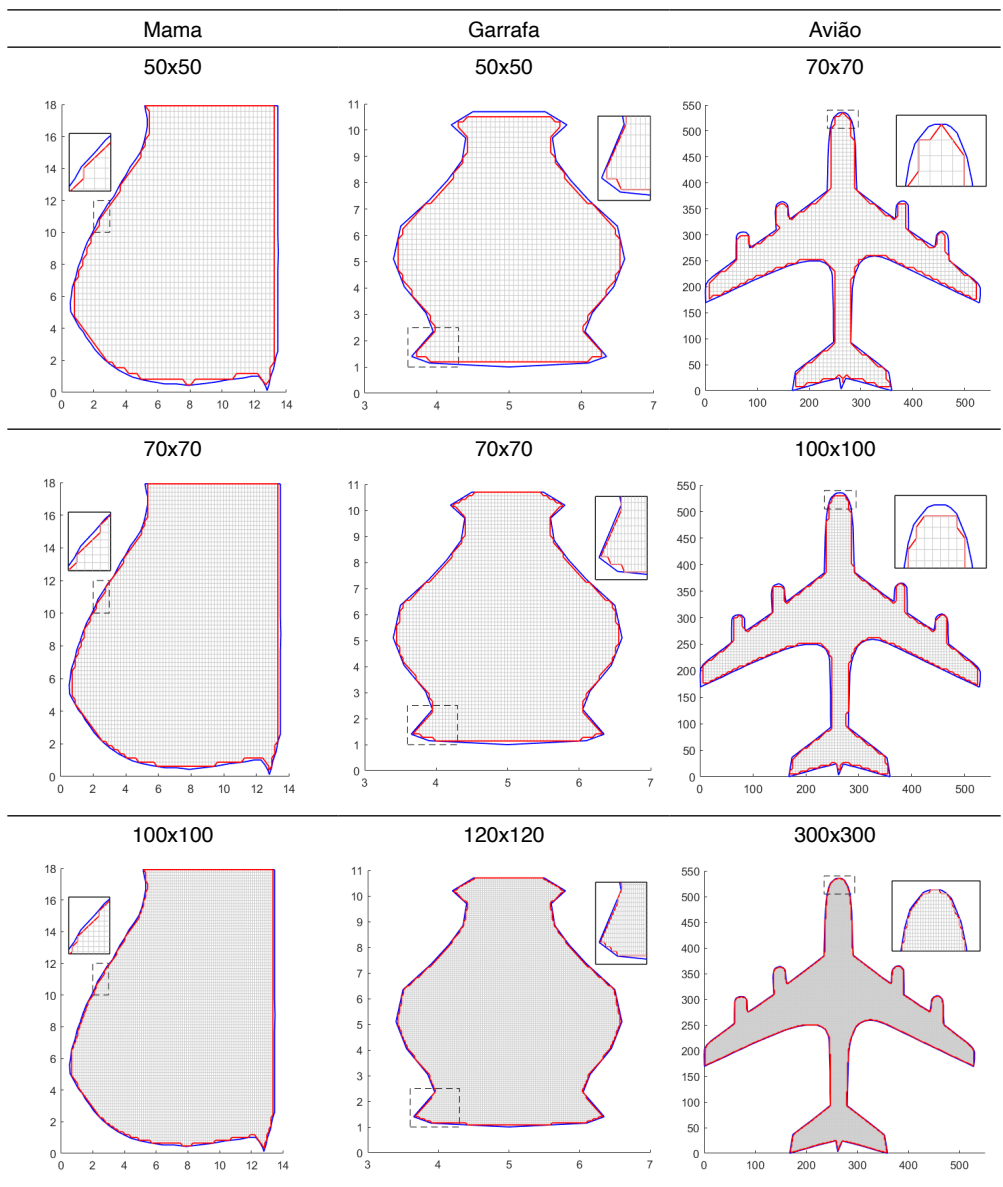

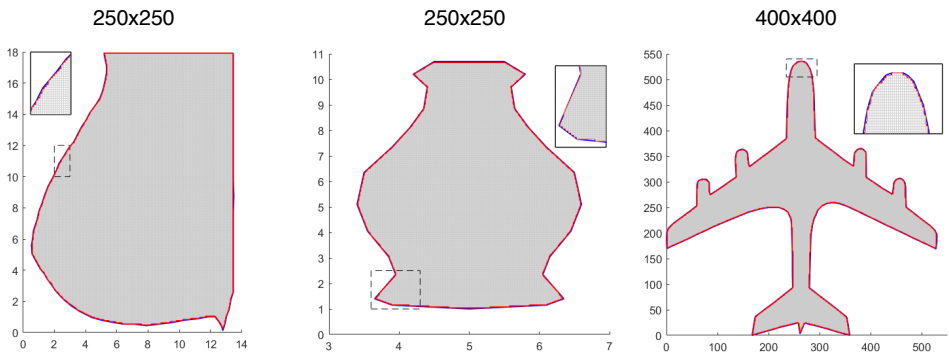

Tabela 2. Resultados obtidos dos contornos dados e aproximados e, malhas considerando as geometrias da mama, garrafa e avião.

Fonte: Os autores.

Verifica-se na Tabelas 1 e 2, como era esperado, que com o refinamento das malhas as diferenças entre as áreas delimitadas pelo contorno dado e aproximado diminuem.

Por fim, as figuras na Tabela 3, apresentam resultados das distâncias entre os nós do contorno aproximado e do contorno dado, para cada ponto da fronteira obtida sobre os nós da malha.

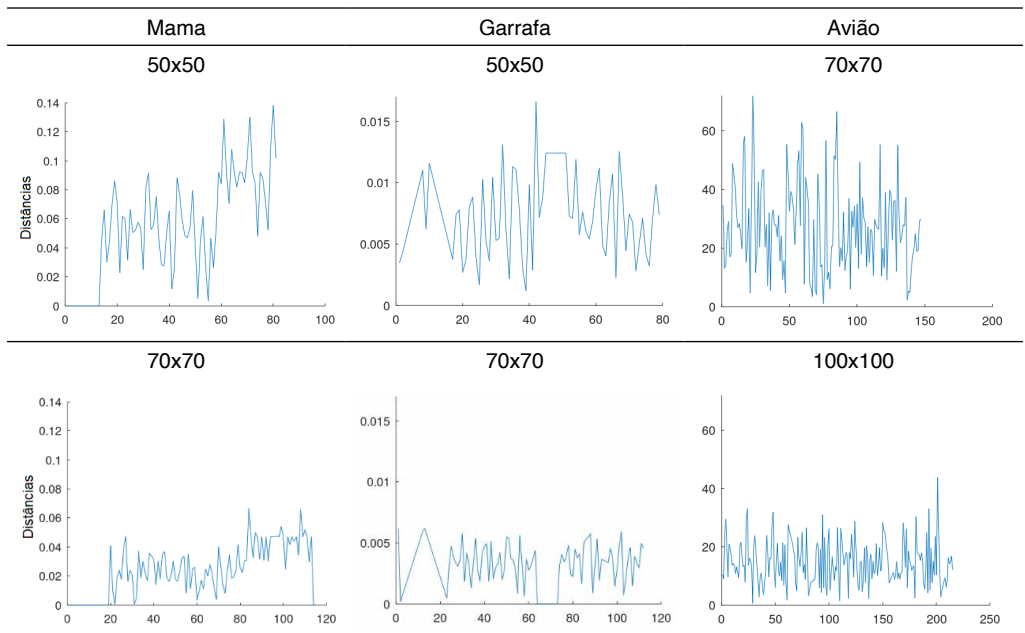

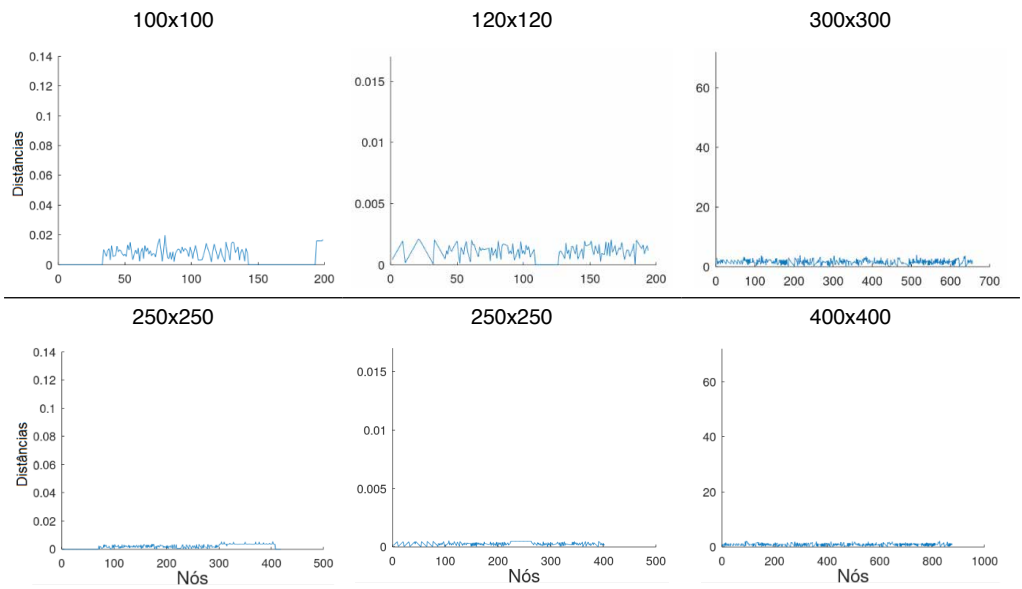

Tabela 3. Resultados das distâncias entre os nós no contorno dado e aproximado, considerando as geometrias da mama, garrafa e avião.

Fonte: Os autores.

Para os resultados apresentados na Tabela 5, optou-se em manter a escala no eixo que descreve as distâncias entre os nós do contorno aproximado e do contorno dado, para uma melhor visualização das diferenças. Observa-se que com o aumento de partições $N_i$ e $N_j$, o número de nós, no contorno aproximado, aumenta significativamente e a diferença entre as distâncias entre os nós do contorno aproximado e do contorno dado diminuem.

Destaca-se que uma das vantagens do algoritmo desenvolvido, refere-se a detectar todos os nós do contorno sobre os pontos de malha, possibilitando o cálculo de diferenças finitas para fins se simulações numéricas de fenômenos como o crescimento tumoral na geometria da mama, a distribuição de calor na geometria da garrafa, análise de escoamento de aeronaves.

## CONCLUSÃO

Desenvolveu-se, neste trabalho, um algoritmo que descreve o contorno aproximado de uma região irregular. O algoritmo resultou em geometrias próximas as geometrias do domínio dado, observou-se que o contorno dado e o contorno aproximado, com o refinamento, tornam-se cada vez mais próximos, comprovando que a diferença entre as áreas delimitadas pelos contornos diminui.

Observou-se que, mesmo utilizando um maior número de nós nas malhas, a geometria do avião apresenta diferenças de áreas maiores em comparação com as outras figuras. Entre os motivos pelos quais isso ocorre, pode-se citar a complexidade da geometria e a grande quantidade de pontos no contorno dado, tornando necessário malhas mais refinadas, para obter diferenças de áreas abaixo de 2%. Além disso, é possível observar que o avião possui região interna mais estreita comparada às outras figuras, fazendo com que as diferenças presentes no contorno se tornem mais relevantes comparadas à área total.

**AGRADECIMENTOS**

**REFERÊNCIAS**

[1] BRADEN, B. The surveyor's area formula. *The College Mathematics Journal,* 17, 4, (1986), 326–337.

[2] CODINA, R., BAIGES, J. Approximate imposition of boundary conditions in immersed boundary methods. *International Journal for Numerical Methods in Engineering 80*, 11 (2009), 1379–1405.

[3] CUMINATO, J. A., AND MENEGUETTE, M. *Discretização de equações diferenciais parciais: técnicas de diferenças finitas*. Sociedade Brasileira de Matemática, 2013.

[4] Eaton, J. W. GNU Octave (version 6.2.0). https://octave.org/doc/v6.2.0/. Acesso em janeiro de 2021.

[5] FERNÁNDEZ-FIDALGO, J., CLAIN, S., RAMÍREZ, L., COLOMINAS, I., NOGUEIRA, X. Very high-order method on immersed curved domains for finite difference schemes with regular cartesian grids. *Computer Methods in Applied Mechanics and Engineering 360* (2020), 112782.

[6] FOUCHER, F., IBRAHIM, M., SAAD, M. Convergence of a positive nonlinear control volume finite element scheme for solving an anisotropic degenerate breast cancer development model. *Computers Mathematics with Applications 76*, 3 (2018), 551 – 578.

[7] FUKUCHI, T. Finite difference method and algebraic polynomial interpolation for numerically solving Poisson's equation over arbitrary domains. *AIP Advances 4*, 6 (2014), 060701.

[8] JOMAA, Z., MACASKILL, C. The embedded finite difference method for the Poisson equation in a domain with an irregular boundary and dirichlet boundary conditions. *Journal of Computational Physics 202*, 2 (2005), 488 – 506.

[9] MAGANIN, J.; ROMEIRO, N. M. L.; CIRILO, E. R.; NATTI, P. L. Simulação de um modelo matemático de crescimento tumoral utilizando diferenças finitas. *Brazilian Journal of Development*, 6, 11, (2020), 87696-87709.

[10] NAOZUKA, G. T. *Geração e análise de qualidade de malhas computacionais em coordenadas curvilíneas*. Dissertação, Universidade Estadual de Londrina – Departamento de Ciência da Computação, Londrina, Pr, Brasil, 2018.

[11] NAOZUKA, G. T.; ROMEIRO, N. M. L.; FELINTO, A. S.; NATTI, P. L.; CIRILO, E. R. Two-dimensional mesh generator and quality analysis of elements on the curvilinear coordinates system. *Semina*: Ciências Exatas e Tecnológicas, 42, 1 (2021), 29–44.

[12] OTHECHAR, P. F. S. *Análise de métodos numéricos de diferenças finitas para solução da equação de Poisson em domínios irregulares*. Dissertação, Programa de Pós-graduação em Matemática Aplicada e Computacional da Universidade Estadual Paulista Júlio de Mesquita Filho, Presidente Prudente, SP, Brasil, 2013.

[13] ROHATGI, A. Webplotdigitizer – versão 4.3. <https://automeris.io/WebPlotDigitizer>, 2020. Acesso em dezembro de 2020.